\documentclass[prl,preprint,superscriptaddress,showpacs,floatfix]{revtex4-1}

\usepackage{latexsym} 
\usepackage{amssymb}  
\usepackage{amsfonts} 
\usepackage{amsbsy}   
\usepackage{amsmath} 
\usepackage{graphicx}       
\usepackage{subfigure}  
\usepackage{multirow}  
\usepackage{bm}   

\renewcommand{\d}{{\rm d}}
\newcommand{\p}{\partial}

\renewcommand{\>}{\rangle} 

\newcommand{\beq}{\begin{equation}}
\newcommand{\eeq}{\end{equation}}
\newcommand{\bea}{\begin{eqnarray}}
\newcommand{\eea}{\end{eqnarray}}
\newcommand{\ba}{\begin{array}}
\newcommand{\ea}{\end{array}}

\newcommand{\e}{{\rm e}}   
\newcommand{\ri}{{\rm i}}

\newcommand{\Tr}{\mbox{Tr}}
\newcommand{\Det}{\mbox{Det}}

\newcommand{\ie}{{i.e.}}
\newcommand{\eg}{{e.g.}}

\makeatletter 

\begin{document}
\title{\bf Stabilizing Non-Hermitian Systems by Periodic Driving}
\author{Jiangbin Gong}
\affiliation{Department of Physics, National University of Singapore, 117542, Singapore}
\affiliation{Centre for Computational Science and Engineering, National University of Singapore, 117542, Singapore}
\author{Qing-hai Wang}
\affiliation{Department of Physics, National University of Singapore, 117542, Singapore}

\date{December 11, 2014}

\begin{abstract}
  The time evolution of a system with a time-dependent non-Hermitian Hamiltonian
  is in general unstable with exponential growth or decay. A periodic driving field may stabilize the dynamics because the eigenphases of the associated Floquet operator may become all real. This possibility can emerge for a {\it continuous} range of system parameters with subtle domain boundaries. It is further shown that the issue of stability of a driven non-Hermitian Rabi model can be mapped onto the band structure problem of a class of lattice Hamiltonians.  As an application, we show how to use the stability of driven non-Hermitian two-level systems (0-dimension in space) to simulate a spectrum analogous to Hofstadter's butterfly that has played a paradigmatic role in quantum Hall physics. The simulation of the band structure of non-Hermitian superlattice potentials with parity-time reversal symmetry is also briefly discussed.
\end{abstract}


\begin{titlepage}
\maketitle
\renewcommand{\thepage}{}          
\end{titlepage}

\section{Introduction}
A number of seminal results have been obtained from studies of periodically driven quantum systems \cite{Casati,HanggiFloquet,Kohler}. One important example, of many decades old and relevant to various research areas (such as cavity quantum electrodynamics), is the coherent Rabi oscillations induced by a driving field \cite{Rabi}. A recent example is the possibility of generating intriguing topological phases by periodic driving \cite{Derek}. Non-perturbative periodic driving is now widely known to be useful in altering symmetry, stability, and topology of a system.  The flexibility in applying a driving field also makes periodically driven systems an attractive platform to realize quantum control and quantum simulation.

Considerable theoretical activities have been devoted to time-independent non-Hermitian systems \cite{BB1,BB2,Review,BerryPT,Longhi,Kottos,mos} that are relevant to optics and to quantum systems with both gain and loss. In particular, time-independent non-Hermitian systems with certain symmetries may still possess a real spectrum before reaching symmetry-breaking points. Experiments on many time-independent non-Hermitian systems were performed \cite{exp1,exp2,Guo,Ruter,exp3,exp4,exp5,exp6,expnew}.  Motivated by these progresses, here we explore periodically driven systems with non-Hermitian Hamiltonians.  Given the vast literature about driven Hermitian systems,  driven non-Hermitian systems are anticipated to be rich and enlightening as well.

The dynamics of a periodically driven system is dictated by its Floquet spectrum. If the Floquet spectrum of a driven non-Hermitian system still falls on the unit circle, the Floquet operator will be unitary up to a similarity transformation.  Then, upon an arbitrary number of driving periods, a Floquet eigenstate only acquires pure phase factors and a general initial state evolves via coherent phase oscillations.  In this manner, periodic driving helps to stabilize the dynamics. As shown below via a non-Hermitian Rabi model, this is feasible (even when the Hamiltonians have complex spectrum during the driving), not just for isolated points in the parameter space, but for a \textit{continuous} range of system parameters.  A previously unknown type of Rabi oscillations, termed ``generalized Rabi oscillations,'' is also found.

Our computational findings are explained through a mapping between a rather general form of driven non-Hermitian two-level systems and the band structure of a class of lattice Hamiltonians.  Depending on the explicit form of the driving, the mapped lattice Hamiltonian can be Hermitian or non-Hermitian. On the one hand, the stability of a driven non-Hermitian problem is connected with a conventional quantum mechanics problem, thus laying a solid foundation for driven non-Hermitian systems. On the other hand, we now have a nonconventional means to simulate lattice Hamiltonians, via a driven two-level system only.   Recognizing the fundamental importance of Hofstadter's butterfly spectrum (HBS) in condensed-matter physics \cite{hofs}, we show how to simulate, in a straightforward manner, the HBS-like spectrum of a class of superlattice Hamiltonians as well as its interesting extensions. Compared with HBS realized in 2-dimensional solid-state materials \cite{Nature1,Nature2,Nature3},
2-dimensional ultracold gases in optical lattices \cite{Bloch,Kett}, and HBS considered in 1-dimensional lattice systems \cite{gongpra, chenshu, 1Dcold, EPJB}, the simulation strategy proposed here is simpler because it is \textit{0-dimensional} in space.

\section{Computational examples}
Let a non-Hermitian but time-periodic dimensionless Hamiltonian be $H(t)=H(t+T)$, where $T$ is the driving period. Throughout we assume scaled and hence dimensionless units (with $\hbar=1$).  The initial time is $t=0$. Unlike previous treatment for time-dependent non-Hermitian systems \cite{GW}, here we stick to the normal Schr\"{o}dinger equation and the conventional Dirac inner product structure.  The time propagator for the period of $[0,t]$ is defined as $U(t)$ and it satisfies
\begin{equation}
\ri \dot{U}(t) = H(t)U(t),
\label{Ueq}
\end{equation}
with the initial condition $U(0)=1$.  Note that the dynamics yielded by Eq.~(\ref{Ueq}) with a time-dependent and non-Hermitian $H(t)$ is nonunitary in general \cite{GW}. The Floquet operator associated with $H(t)$
is given by $U(T)$, with its spectrum determined by the eigenvalue equation
\begin{equation}
U(T)|\phi_n\> = \e^{\ri \beta_n} |\phi_n\>,
\end{equation}
where the $n$th Floquet eigenstate is $|\phi_n\>$ with the eigenvalue  $\e^{\ri \beta_n}$.  Of particular interest is the situation when all $\beta_n$ are indeed real and hence $\e^{\ri \beta_n}$ are pure phase factors. If this is true, then
\begin{equation}
U(T) = S D S^{-1},
\label{generalU}
\end{equation}
where $D$ is a diagonal unitary matrix with   phase factors $\e^{\ri  \beta_n}$ on the diagonal and $S$ is a
similarity transformation. A Floquet operator satisfying Eq.~(\ref{generalU}) is said to possess ``extended unitarity,'' which then yields $U^{N}(T)=SD^{N}S^{-1}$. Thus, if extended unitarity emerges, then only pure phase factors $\e^{\ri N\beta_n}$ enter into the time evolution operator for (arbitrary) $N$ driving periods.  The dynamics is hence stable because there is no exponential growth or decay with $N$.

To make a driven non-Hermitian system as simple as possible, one may introduce non-Hermitian terms to a two-level Rabi model, which is much relevant to understanding the evolution of two optical polarizations in a nontransparent medium \cite{BerryPT}.
We discuss two specific examples, characterized by two real parameters $\gamma$ and $\mu$  with $T=1$. In the first example, we choose $H_1(t)=\gamma \sigma_z + \ri \mu[ \cos(2\pi t) +\sin(4\pi t)]\sigma_x,$ where $\sigma_x$ and $\sigma_z$ are the standard Pauli matrices.  The driving component of $H_1(t)$ is anti-Hermitian, with two driving frequencies $2\pi$ and $4\pi$ (this is to indicate  a rather arbitrary periodic driving).  The findings are summarized in Fig.~1(a), obtained by carefully scanning the values of $\gamma$ and $\mu$ and then checking the Floquet spectrum. There the shaded regimes represent the domains of extended unitarity.  One might na\"{i}vely think that extended unitarity only occurs accidently. Contrary to this intuition, it is seen to emerge for a wide and continuous range of $\gamma$ and $\mu$, with highly intricate domain boundaries.  It should also be stressed that domain of extended unitarity is \textit{not at all} the domain for $H_1(t)$ to have a real instantaneous spectrum.
\begin{figure}[h!]
$\begin{array}{cc}
\quad~~\text{(a)}& \quad~~\text{(b)}\\
\includegraphics[width=0.48\columnwidth]{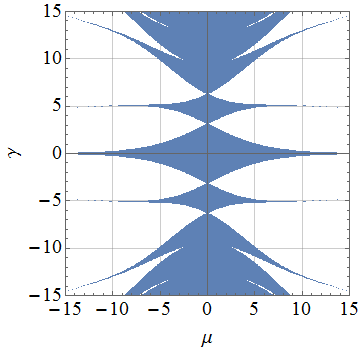}&
\includegraphics[width=0.48\columnwidth]{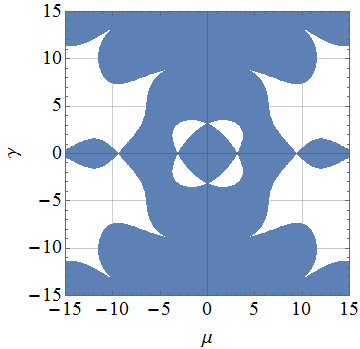}
\end{array}$
\caption{\label{fig:2x2gm}(color online) Phase diagrams for two non-Hermitian extensions of the Rabi model, with Hamiltonians $H_1(t)$ (a) and $H_2(t)$ (b) defined in the text. Shaded regimes represent extended unitarity and hence stabilization
afforded by periodic driving.}
\end{figure}

Let us turn to the second example with the Hamiltonian
$H_2(t)= \gamma \sigma_z + \ri\mu[\sin(2\pi t)+\ri]\sigma_x$.   The static component of $H_2$ now has a component
parallel to the non-Hermitian driving term. Extended unitarity also emerges, with the corresponding phase diagram in Fig.~1(b) displaying again subtle boundaries. Note that the instantaneous eigenvalues of $H_2(t)$ are not real except $t/T=0,\ 1/2,\ 1$. That is, stabilization is possible, even when the instantaneous spectrum of $H_2(t)$ is complex during almost the entire period of driving. In principle, one is allowed to introduce and then scan over more system parameters other than $(\mu,\gamma)$ or scan $(\mu,\gamma)$ in the complex domain. However, it would be challenging to present a high-dimensional phase diagram.  Qualitatively similar stabilization is also observed in many other non-Hermitian variants of the Rabi model, including those with a Hermitian driving but non-Hermitian static component.

\begin{figure}[h!]
$\begin{array}{cc}
\quad~~\text{(a)}&\quad~~\text{(b)}\\
\includegraphics[width=0.48\columnwidth]{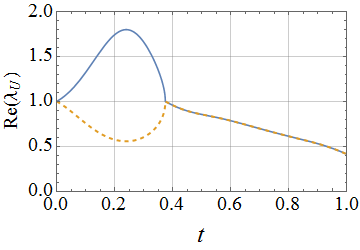}&
\includegraphics[width=0.48\columnwidth]{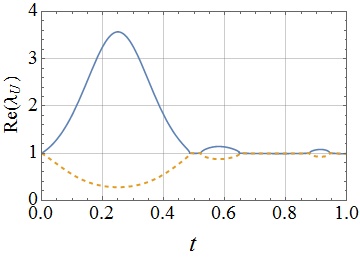}\\
\quad~~\text{(c)}&\quad~~\text{(d)}\\
\includegraphics[width=0.48\columnwidth]{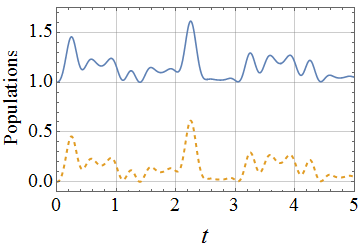}&
\includegraphics[width=0.48\columnwidth]{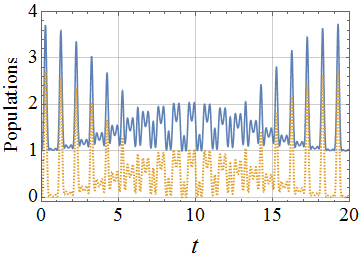}
\end{array}$
\caption{\label{fig:2x2Rabi} (color online)  Top panels depict the real part of the eigenvalues of $U(t)$, denoted $\lambda_U$, during one period of driving. Bottom panels show generalized Rabi oscillations via populations of spin up (blue lines) and spin down (orange dashed lines). The initial state is up and the Hamiltonian is $H_1(t)$ defined in the text.
In (a) and (c) $\gamma = 1$ and $\mu = 2$, in (b) and (d) $\gamma = 0.1$ and $\mu = 4$.}
\end{figure}

Next we take two sets of $(\mu,\gamma)$ from the domain of extended unitarity of $H_1$ to further digest the dynamics.  Firstly, we analyze in Fig.~\ref{fig:2x2Rabi} the real part of the spectrum of $U(t)$ for $t\in [0,T]$.  Because $H_1$ is traceless, it can be shown that if and only the two eigenvalues of $U(t)$ have the same real parts, then the eigenvalues of $U(t)$ can be written as $\exp (\pm \ri \beta)$ with a real $\beta$ \cite{note}. The top two panels of Fig.~\ref{fig:2x2Rabi} depict the splitting of one common real part of the eigenvalues into two, followed by a recombination of two into one. Such splitting and recombination behavior may occur several times within one period.   This vividly shows that, at times not equal to multiple periods of $T$, $U(t)$ does not necessarily have real eigenphases. Thus, yielding extended unitarity (at $t=NT$) still allows for rather complicated and potentially exotic dynamics within one driving period.
Secondly, let us examine the population dynamics in the presence of extended unitarity. The initial state is assumed to be the first state and the corresponding results are shown in the bottom panels of Fig.~\ref{fig:2x2Rabi}.  Stable and coherent population oscillations are observed in Fig.~\ref{fig:2x2Rabi}(c) and  Fig.~\ref{fig:2x2Rabi}(d), representing a type of generalized Rabi oscillations. Interestingly, the total population on the two states may go beyond unity, which reminds us that the system dynamics is stable but not unitary.  A careful check further shows that in the two shown examples the population difference (rather than the population sum) is unity at all times. This is because the driving field happens to be perpendicular to the static field, whose direction is also the direction of population measurement.

\section{Mapping stability to band structure problems}
To gain insights into why stability can be thus restored, we now consider a class of traceless and non-Hermitian two-level Hamiltonians subject to one-parameter periodic modulation:
\begin{equation}
H(t)= [a {\bf n}_3 + \ri b(t) {\bf n}_1]\cdot{\bm \sigma},
\end{equation}
where $a$ is time-independent and $b(t)=b(t+T)$ is a complex periodic variable, ${\bm \sigma}=(\sigma_x,\sigma_y,\sigma_z)$, and $\{{\bf n}_1, {\bf n}_2, {\bf n}_3\}$ are an arbitrary but fixed set of vectors forming a right-handed basis set.
For reasons to be elaborated below, $a^2$ is assumed to be real. We next expand $U(t)$ in the same representation, yielding
\begin{equation}
U(t)=u_0(t) + \sum_{i=1}^{3} u_i(t) {\bf n}_i\cdot {\bm \sigma},
\label{Uexp}
\end{equation}
with complex expansion coefficients $u_i(t)$ under the initial conditions $u_0(0)=1$, and $u_i(0)=0$ for $i=1,2,3$. The two eigenvalues of $U(T)$ are hence given by $\e^{\pm \ri \beta}= u_0(T) \pm \ri\sqrt{1-u_0^2(T)}$. Clearly then, for $\beta$ to be a real phase (hence extended unitarity), it is sufficient and necessary for $u_0(T)$ to be real, with $-1\leq u_0(T)\leq 1$, such that $\beta=\arccos[u_0(T)]$.  Because the eigenvalues of $U(NT)$ are simply $\e^{\pm  \ri N\beta}$,  this condition also leads to $u_0(NT)=\cos(N\beta)$ and hence $-1\leq u_0(NT)\leq 1$ for arbitrary $N$.

To proceed we substitute Eq.~(\ref{Uexp}) into Eq.~(\ref{Ueq}), yielding
\begin{equation}
\left\{
\begin{array}{rcl}
\dot{u}_0(t) &=& b(t) u_1(t) -\ri a u_3(t)\\
\dot{u}_1(t) &=& b(t) u_0(t) -a u_2(t)\\
\dot{u}_2(t) &=& a u_1(t) -\ri b(t) u_3(t)\\
\dot{u}_3(t) &=& -\ri a u_0(t) +\ri b(t) u_2(t)\;.
\end{array}
\right.
\label{eqn:U2x21st}
\end{equation}
Differentiating Eq.~(\ref{eqn:U2x21st}) again and canceling the first order derivatives, we obtain the equations satisfied by $u_0(t)$ and $u_1(t)$,
\begin{equation}
\left[-  \frac{\d^2}{\d t^2} + \left(
\begin{array}{cc}
b^2(t) & \dot{b}(t)\\
\dot{b}(t) & b^2(t)\\
\end{array}
\right) \right] \left(
\begin{array}{c}
u_0(t)\\
u_1(t)
\end{array}\right)
=a^2 \left(
\begin{array}{c}
u_0(t)\\
u_1(t)
\end{array}\right).
\label{u0u1eq}
\end{equation}
By mapping $t$ in Eq.~(\ref{u0u1eq}) onto a space variable $x$ and introducing $\psi^{\pm}(x) = u_0(x)\pm u_1(x)$, we turn Eq.~(\ref{u0u1eq}) into two decoupled stationary Schr\"{o}dinger equations
\begin{equation}
-  \frac{\d^2}{\d x^2}{\psi}^{\pm}(x) + \left[b^2(x) \pm \frac{\d b(x)}{\d x} \right] \psi^{\pm}(x) = a^2 \psi^{\pm}(x),
\label{eqv}
\end{equation}
which describe a particle of mass $1/2$ moving in one of the two periodic potentials
\begin{equation}
V^{\pm}(x)\equiv b^2(x) \pm \frac{\d b(x)}{\d x}
\end{equation}
of lattice constant $T$, \ie, $V^{\pm}(x+T)= V^{\pm}(x)$. For example, if $b(t)$ is a time-periodic square function,
then $V^{\pm}(x)$ will become a generalized Dirac-Kronig-Penney model as it is comprised by $\delta$ potentials with alternating signs.  Interestingly, because $V^{+}(x)$ and $V^{-}(x)$ naturally form a super-symmetric potential pair, they yield identical spectrum \cite{supersymmetry-note}. As such, nontrivial solutions to Eq.~(\ref{eqv}) with a common eigenvalue $a^2$ should exist for both $\psi^{+}(x)$ and $\psi^{-}(x)$. This being the case, one may just focus only on $\psi^{+}(x)$ and $V^{+}(x)$ in Eq.~(\ref{eqv}).  Due to this mapping, below we do not clearly distinguish between $t$ and $x$ variables when the context is clear.

The extended unitarity condition  $-1\leq u_0(NT)\leq 1$ can now be better digested. Intuitively, if there is a complex eigenphase $\beta$,  then the condition $-1\leq u_0(NT)\leq 1$ is violated because of an exponential growth of $u_0(NT)$ vs $N$.  But if this exponential growth occurs, then $\psi^{\pm}(NT)= u_0(NT)\pm u_1(NT)$ diverges with $N$ and hence cannot be a Bloch eigenfunction at energy $a^2$ [see Eq.~(\ref{eqv})].  That is to say, in order to achieve dynamics stabilization,  the system parameters must be chosen such that the driving field profile $b(t)$ admits Bloch wavefunctions at energy eigenvalue $a^2$ [via Eq.~(\ref{eqv})].  On the other hand, if $\psi^{\pm}(x)$ is a Bloch wavefunction, then one can indeed explicitly construct from  $\psi^{\pm}(x)$ a solution of $u_0(t)$ satisfying $-1\leq u_0(NT)\leq 1$ and its initial condition.
We have thus identified a mapping between the band structure of $V^{\pm}(x)$ and the stability in a driven non-Hermitian system. More detailed analysis shows that through this general mapping, the real eigenphase $\beta$ of $U(T)$ becomes the Bloch quasi-momentum times the lattice period in the band structure problem for $\psi^{\pm}(x)$.

\begin{figure}[h!]
$\begin{array}{cc}
\quad~~\text{(a)} &\quad~~\text{(b)}\\
\includegraphics[width=0.48\columnwidth]{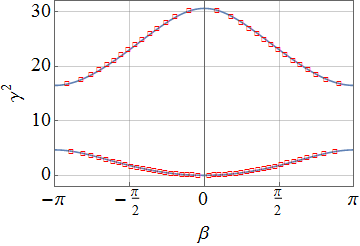}&
\includegraphics[width=0.48\columnwidth]{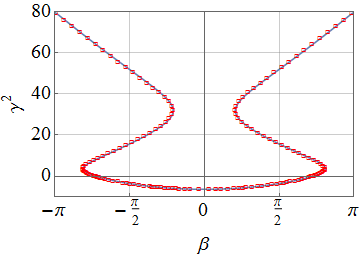}
\end{array}$
\caption{\label{fig:dispersion-relation}(color online) Identical dispersion relations obtained by direct band-structure calculations using $V_1^+$ or $V_2^+$ defined in the text (blue lines) or by checking if extended unitarity occurs (red squares).  Panel (a) is for $H_1$ with $\mu=2$ and panel (b) is for $H_2$ with $\mu=4$. The negative $\gamma^2$ part in (b) is obtained by scanning the parameter $\gamma$ in the purely imaginary domain. Results here confirm our theoretical mapping.}
\end{figure}

Returning to the case of $H_1(t)$, ${\bf  n}_1=\hat{x}$, ${\bf  n}_3=\hat{z}$, we have $a=\gamma$, $b(t)=\mu[\sin(2\pi t)+\cos(4\pi t)]$.   Then the mapped lattice potential becomes $V_1^+(x)=\mu^2 [\sin^2(2\pi x) + \cos^2(4\pi x) +2\sin(2\pi x)\cos(4\pi x)]+2\pi\mu [\cos(2\pi x)-2\sin(4\pi x)]$, a real superlattice potential.  In the same manner, $H_2(t)$ is mapped to a lattice potential $V_2^+(x)=\mu^2[\sin^2(2\pi x)-1+2\ri\sin(2\pi x)]+2\pi \mu \cos(2\pi x)$. $V_2^+(x)$ is seen to be complex, but it is invariant upon a joint time-reversal and parity (${\cal PT}$) operation, the so-called ${\cal PT}$ invariance.  The possibility of having real spectrum ($a^2$ is constructed to be real) under ${\cal PT}$ symmetry ensures that we still possibly have Bloch wavefunctions for $V_2^+(x)$. Therefore, for both examples of $H_1(t)$ and $H_2(t)$, the phase diagrams in Fig.~1 can now be understood as the collection of all possible real  band energy eigenvalues $a^2=\gamma^2$ as a function of a second system parameter $\mu$.  The origin of the boundaries seen in Fig.~1 is hence identified as the presence of energy gaps for the real potential $V_1^+(x)$, or as an interplay of band gaps and symmetry-breaking points for the ${\cal PT}$-symmetric complex potential $V_2^+(x)$. To further check our understandings, for one value of $\mu$ we record $\beta$ when extended unitarity occurs and then plot $\gamma^2$ vs $\beta$ for $H_1$ and $H_2$ (red squares). The results are then compared in Fig.~3 with band dispersion relations obtained from direct band-structure calculations for $V_1^+$ or  $V_2^+$.  The agreement seen in Fig.~3 confirms our exact mapping described above.

\section{Quantum simulation}
\begin{figure}[h!]
$\begin{array}{cc}
\quad~~\text{(a)} &\quad~~\text{(b)}\\
\includegraphics[width=0.48\columnwidth]{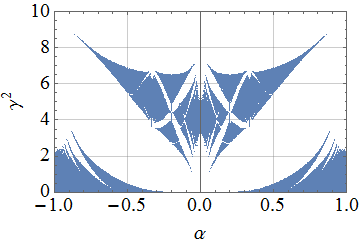}&
\includegraphics[width=0.48\columnwidth]{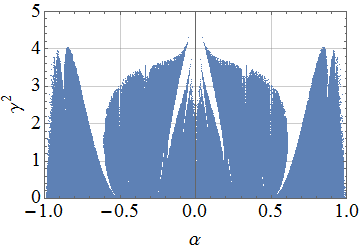}
\end{array}$
\caption{(color online) Simulation of a phase pattern analogous to Hofstadter's butterfly spectrum (HBS) (a) and
an extension of HBS for non-Hermitian but ${\cal PT}$-symmetric Hamiltonians (b), using
driven non-Hermitian Hamiltonians $H_3(t)$ (a) and $H_4(t)$ (b) defined in the text with both $\mu=2$.
Blue dots represent extended unitarity restored by periodic driving.}
\label{fig:butterfly}
\end{figure}

To motivate potential experimental interest,  let us now investigate two non-Hermitian Rabi models upon introducing a parameter $\alpha$ that describes the period ratio of two commensurable driving periods.  Consider first $H_3(t)= \gamma \sigma_z + \ri \mu[ \cos(2\pi t) +\cos (2 \alpha \pi t)]\sigma_x$. If  the parameter $\alpha$ is a  rational number with $\alpha=p/q$ ($p,q$ two co-prime integers), the mapped superlattice potential $V_3^+$, comprised by a base lattice of period unity and additional superlattice components, still has a period $q$. The $\alpha$ parameter hence resembles the role of the magnetic flux per plaque in the HBS Hamiltonian of the original quantum Hall problem \cite{hofs}. For a fixed value of $\mu$, we obtain the phase diagram of extended unitarity in terms of $\gamma^2$ vs a varying rational $\alpha$. The results are shown in Fig.~\ref{fig:butterfly}(a).   The shown phase diagram of extended unitarity is indeed
highly similar to HBS. In particular, many clear gaps and intriguing domain boundary profiles are found.  This is achieved without the use of a magnetic field, a clean 2-dimensional material,  or even a 1-dimensional lattice potential.
Given the paradigmatic role of HBS in understanding quantum Hall physics \cite{hofs}, our findings in Fig.~\ref{fig:butterfly}(a) have paved a nonconventional way towards the simulation of quantum Hall physics, including topological phase transitions.  For example, it will be valuable to examine the topological characterizations and implications of the gaps seen in Fig.~\ref{fig:butterfly}(a).

Next we consider $H_4(t)= \gamma \sigma_z + \ri \mu[\ri\cos(2\pi t) +\sin (2 \alpha \pi t)]\sigma_x$.  In this case, the associated superlattice potential $V_4^+(x)$ is non-Hermitian but is apparently ${\cal PT}$-symmetric.  This situation hence represents a complex extension of the original HBS problem.  Interestingly, the resulting phase diagram of $H_4$ shown in Fig.~\ref{fig:butterfly}(b) has much less gaps and is thus quite different from a conventional HBS.  Upon a careful inspection, the lack of many gaps here is found to be connected with ${\cal PT}$-symmetry breaking in the mapped problem. That is,  a gap often closes at the critical quasi-momentum value for which the spectrum of a ${\cal PT}$-symmetric lattice becomes complex [see Fig.~\ref{fig:dispersion-relation}(b)].

\section{Concluding remarks}
Stabilization by periodic driving does not need the instantaneous spectrum of the Hamiltonian to be real. This further extends opportunities in studies of non-Hermitian systems. Stabilization can be also achieved in other non-Hermitian systems with more levels. Extended unitarity realized by periodic driving is expected to be more useful than what is learned here.  We also note a recent stimulating study \cite{PRAwork}, where the emphasis is placed on how ${\cal PT}$-symmetry of driven systems is broken by a close-to-resonance perturbation. The mapping found here can be used to explain some results in Ref.~\cite{PRAwork}.

\begin{acknowledgments}
{\bf Acknowledgments:}
QhW would like to thank Prof.~Weixiao Shen for enlightening comments regarding the highly complex nature of driven non-Hermitian systems.
\end{acknowledgments}



\begin{thebibliography}{99}
\bibitem{Casati}
    G.~Casati and B.V.~Chirikov, \textit{Quantum Chaos: Between Order and Disorder} (Cambridge University Press, New York, 1995).
\bibitem{HanggiFloquet}
    M.~Grifoni and P.~H\"{a}nggi, Phys.~Rep.~{\bf 304}, 229 (1998).
\bibitem{Kohler}
    S.~Kohler, J.~Lehmann, and P.~H\"{a}nggi, Phys.~Rep.~{\bf 406}, 379 (2005).
\bibitem{Rabi}
    I.I.~Rabi, Phys.~Rev.~{\bf 49}, 324 (1937); \textit{ibid}~{\bf 51}, 652 (1937).
\bibitem{Derek}
    For a rather comprehensive list of recent studies of periodically driven systems in the context of toplogical phases of matter, see \eg,  D.Y.H.~Ho and J.B.~Gong, \prb~{\bf 90}, 195419 (2014).
\bibitem{BB1}
    C.M.~Bender and S.~Boettcher, \prl~{\bf 80}, 5243 (1998).
\bibitem{BB2}
    C.M.~Bender, D.C.~Brody, and H.F.~Jones, \prl~{\bf 89}, 270401 (2002).
\bibitem{Review}
    C.M.~Bender, Rep.~Prog.~Phys.~{\bf 70}, 947 (2007).
\bibitem{BerryPT}
    M.V.~Berry, J.~Opt.~{\bf 13}, 115701 (2011).
\bibitem{Longhi}
    S.~Longhi, \prl~{\bf 103}, 123601 (2009).
\bibitem{Kottos}
    C.T.~West, T.~Kottos, and T.~Prosen, \prl~{\bf 104}, 054102 (2010).
\bibitem{mos}
    A.~Mostafazadeh, J.~Math.~Phys.~{\bf 43}, 205 (2002); \textit{ibid}~{\bf 43}, 2814 (2002).
\bibitem{exp1}
    Z.H.~Musslimani, K.G.~Makris, R.~El-Ganainy, and  D.N.~Christodoulides, \prl~{\bf 100}, 030402 (2008).
\bibitem{exp2}
    K.G.~Makris, R.~El-Ganainy, D.N.~Christodoulides, and Z.H.~Musslimani, \prl~{\bf 100}, 103904 (2008).
\bibitem{Guo}
    A.~Guo, G.J.~Salamo, D.~Duchesne, R.~Morandotti, M.~Volatier-Ravat, V.~Aimez, G.A.~Siviloglou, and D.N.~Christodoulides, \prl~{\bf 103}, 093902 (2009).
\bibitem{Ruter}
    C.E.~R\"{u}ter, K.G.~Makris, R.~El-Ganainy, D.N.~Christodoulides, M.~Segev, and D.~Kip, Nature Physics {\bf 6}, 192 (2010).
\bibitem{exp3}
    J.~Schindler, A.~Li, M.C.~Zheng, F.M.~Ellis, and T.~Kottos, \pra~{\bf 84}, 040101 (2011).
\bibitem{exp4}
    A.~Regensburger, C.~Bersch, M.A.~Miri, G.~Onishchukov, D.N.~Christodoulides, and U.~Peschel, Nature {\bf 488}, 167 (2012).
\bibitem{exp5}
    A.~Regensburger, M.A.~Miri, C.~Bersch, J.~N\"{a}ger, G.~Onishchukov, D.N.~Christodoulides, and U.~Peschel, \prl~{\bf 110}, 223902 (2013).
\bibitem{exp6}
    M.~Brandstetter, M.~Liertzer, C.~Deutsch, P.~Klang, J.~Sch\"{o}berl, H.E.~T\"{u}reci, G.~Strasser,
    K.~Unterrainer, and S.~Rotter, Nature Commu. {\bf 5}, 4034 (2014).
\bibitem{expnew}
    B.~Peng, S.K.~Ozdemir, S.~Rotter, H.~Yilmaz, M.~Liertzer, F.~Monifi, C.M.~Bender, F.~Nori, and L.~Yang, Science, \textbf{346}, 328 (2014).
\bibitem{hofs}
    D.R.~Hofstadter, \prb~{\bf 14}, 2239 (1976).
\bibitem{Nature1}
    C.R.~Dean \textit{et al.}, Nature {\bf 497}, 598 (2013).
\bibitem{Nature2}
    L.A.~Ponomarenko \textit{et al.}, Nature {\bf 497}, 594 (2013).
\bibitem{Nature3}
   B.~Hunt \textit{et al.}, Science {\bf 340}, 1427 (2013).
\bibitem{Bloch}
    M.~Aidelsburger, M.~Atala, M.~Lohse, J.T.~Barreiro, B.~Paredes, and I.~Bloch, \prl~{\bf 111}, 185301 (2013).
\bibitem{Kett}
    H.~Miyake, G.A.~Siviloglou, C.J.~Kennedy, W.C.~Burton, and W.~Ketterle, \prl~{\bf 111}, 185302 (2013).
\bibitem{gongpra}
    J.~Wang and J.B.~Gong, \pra~{\bf 77}, 031405 (2008).
\bibitem{chenshu}
    L.J.~Lang, X.~Cai, and S.~Chen, \prl~{\bf 108}, 220401 (2012).
\bibitem{1Dcold}
    A.~Celi, P.~Massignan, J.~Ruseckas, N.~Goldman, I.B.~Spielman, G.~Juzeliunas, M.~Lewenstein, \prl~{\bf 112}, 043001 (2014).
\bibitem{EPJB}
    L.W.~Zhou, H.L.~Wang, D.Y.H.~Ho and J.B.~Gong, EPJB {\bf 87}, 204 (2014).
\bibitem{GW}
    J.B.~Gong and Q.-h.~Wang, \pra~{\bf 82}, 012103 (2010); J.~Phys.~A {\bf 46}, 485302 (2013).
\bibitem{note}
    For a traceless Hamiltonian $H(t)$,  Liouville's formula $\ri \frac{\p}{\p t} \Det[U(t)] = \Tr [H(t)] \Det[U(t)]$ indicates that $\Det[U(t)]$, the determinant of its propagator $U(t)$ must stay at unity. That two eigenvalues of a two-dimensional $U$ have the same real parts then becomes equivalent to real eigenphases, \ie, $\exp(\pm \ri\beta)$ with a real $\beta$.
\bibitem{supersymmetry-note}
    That a supersymmetry pair generates identical spectrum is often discussed under Dirichlet boundary condition. It is nevertheless found to be also true here. For supersymmetric mechanics, see the review article F.~Cooper, A.~Khare, and U.~Sukhatme, Phys.~Rep.~\textbf{251} 267 (1995) and the referemces therein.
\bibitem{PRAwork}
    Y.N.~Joglekar, R.~Marathe, P.~Durganandini, and R.K.~Pathak, \pra~{\bf 90}, 040101 (2014).
\end{thebibliography}
\end{document}